\documentclass[twocolumn,showpacs,preprintnumbers,prd,nofootinbib]{revtex4}
\usepackage{epsfig,amsmath,amssymb,verbatim,colordvi,color}
\usepackage[bookmarks=false]{hyperref}

\newcommand{\beq}{\begin{equation}}
\newcommand{\eeq}{\end{equation}}
\newcommand{\bea}{\begin{eqnarray}}
\newcommand{\eea}{\end{eqnarray}}
\newcommand{\Fig}[1]{Fig.\,\ref{#1}}

\newcommand{\eq}[1]{(\ref{#1})}

\newcommand{\Sec}[1]{Sec.\,\ref{#1}}

\newcommand{\f}{\frac}

\newcommand{\gs}{g}
\newcommand{\as}{\alpha_s}
\newcommand{\aem}{\alpha}

\newcommand{\mc}{m_c}
\newcommand{\mb}{m_b}
\newcommand{\ms}{m_s}

\newcommand{\mub}{\mu_b}

\newcommand{\NL}{N_L}

\newcommand{\TR}{T_F}

\newcommand{\GeV}{{\rm GeV}}

\newcommand{\MSbar}{\overline{\rm MS}}

\newcommand{\ord}{{\cal O}}
\def\unit{\leavevmode\hbox{\small1\kern-3.6pt\normalsize1}}
\newcommand{\eps}{\epsilon}

\newcommand{\BXsga}{\bar{B} \to X_s \gamma}

\newcommand{\BRga}{{\cal B} (\BXsga)}

\newcommand{\BRXc}{{\cal B} (\bar{B} \to X_c \ell \bar{\nu})}

\newcommand{\etal}{{\it et al.}}
\newcommand{\ie}{{\it i.e.}}

\let\oldmarginpar\marginpar
\renewcommand\marginpar[1]{\-\oldmarginpar[\raggedleft\scriptsize\sf
#1]{\raggedright\scriptsize\sf #1}} 

\begin{document}

\preprint{MZ-TH/10-30} 

\title{\boldmath Chromomagnetic Dipole-Operator Corrections in
  $\BXsga$ at $\ord (\beta_0 \hspace{0.15mm} \as^2)$ \unboldmath}

\author{Andrea~Ferroglia$^{1}$ and Ulrich~Haisch$^{2}$} 

\affiliation{
$^1$Physics Department, 
   New York City College of Technology, 
   300 Jay Street,
   Brooklyn New York 11201, USA \\
\vspace{.3cm}
$^2$Institut f\"ur Physik (THEP),
    Johannes Gutenberg-Universit\"at,
    D-55099 Mainz, Germany 
}

\date{\today}

\begin{abstract}
  \noindent 
  We calculate the fermionic corrections to the photon-energy spectrum
  of $\BXsga$ which arise from the self-interference of the
  chromomagnetic dipole operator $Q_8$ at next-to-next-to-leading
  order by applying naive non-abelianization. The resulting ${\cal O}
  (\beta_0 \hspace{0.15mm} \as^2)$ correction to the $\BXsga$
  branching ratio amounts to a relative shift of $+0.12\%$ ($+0.27\%$)
  for a photon-energy cut of $1.6 \, {\rm GeV}$ ($1.0 \, {\rm
    GeV}$). We also comment on the potential size of resummation and
  non-perturbative effects related to $Q_8$.
\end{abstract}


\maketitle

\section{Introduction} 
\label{sec:intro}

The inclusive radiative $B$-meson decay $\BXsga$ represents the
``standard candle'' of quark-flavor physics. It tests the electroweak
structure of the underlying theory and provides information on the
couplings and masses of heavy virtual particles appearing as
intermediate states in and beyond the Standard Model (SM). See
\cite{Haisch:2008ar} for a concise overview.

The present experimental world average for a photon-energy cut of
$E_\gamma > E_0$ with $E_0 = 1.6 \, \GeV$ in the $\bar{B}$-meson rest
frame reads \cite{Barberio:2008fa}
\bea \label{eq:WA} 
\BRga_{\rm exp}^{E_\gamma > 1.6 \, {\rm GeV}} = \left ( 3.55 \pm 0.24
  \pm 0.09 \right ) \cdot 10^{-4} . \hspace{1.25mm}
\eea
The quoted value includes various measurements from CLEO, BaBar, and
Belle \cite{various} and has a total error of below $8 \%$, which
consists of a combined statistical and systematic error as well as a
systematic uncertainty due to the shape function.

In order to make full use of the available data, the SM calculation of
$\BXsga$ should be performed with similar or better precision. This
goal can only be achieved with dedicated calculations of
next-to-next-to-leading order (NNLO) QCD effects in
renormalization-group improved perturbation theory. Considerable
effort has gone into such computations. The necessary two- and
three-loop matching was performed in \cite{Bobeth:1999mk} and
\cite{Misiak:2004ew}, while the mixing at three and four loops was
calculated in \cite{3mix} and \cite{Czakon:2006ss}. The two-loop
matrix element including bremsstrahlungs corrections of the photonic
dipole operator $Q_7$ was found in \cite{1stQ7}, confirmed in
\cite{2ndQ7}, and extended to include the full charm-quark mass
dependence in \cite{Asatrian:2006rq}. The three-loop matrix elements
of the current-current operators $Q_{1,2}$ were derived in
\cite{Bieri:2003ue} within the large-$\beta_0$ approximation. A
calculation that goes beyond this approximation employs an
interpolation in the charm-quark mass \cite{Misiak:2006ab}.
Contributions involving a massive quark-loop insertion into the gluon
propagator of the three-loop $Q_{1,2}$ matrix elements are also known
\cite{Boughezal:2007ny}. Calculations of other missing NNLO pieces,
such as the $(Q_7, Q_8)$ interference were recently completed
\cite{Q7Q8}. Further details on the status of the NNLO corrections to
the branching ratio of $\BXsga$ can be found in \cite{NNLO}.

Combining the results listed above, it was possible to obtain the
first estimate of the $\BXsga$ branching ratio at NNLO. For $E_0 = 1.6
\, \GeV$ the result of the improved SM evaluation is given by
\cite{Misiak:2006zs, Misiak:2006ab}\footnote{Several NNLO corrections
  (see \cite{Asatrian:2006rq, Boughezal:2007ny, Q7Q8} and partly
  \cite{Czakon:2006ss}) that were calculated after the publication of
  \cite{Misiak:2006zs, Misiak:2006ab} are not included in the central
  value of (\ref{eq:WA}), but remain within the quoted perturbative
  higher-order uncertainty of $3\%$.}
\beq \label{eq:SM} 
\BRga_{\rm SM}^{E_\gamma > 1.6 \, {\rm GeV}} = (3.15 \pm 0.23) \cdot
10^{-4} \, ,
\eeq 
where the individual uncertainties from non-perturbative corrections
($5 \%$), parametric dependences ($3 \%$), higher-order perturbative
effects $(3 \%)$, and the interpolation in the charm-quark mass ($3
\%$) have been added in quadrature to obtain the total error. More
details on the phenomenological NNLO analysis including the list of
input parameters can be found in \cite{Misiak:2006ab}. A systematic
study of hadronic effects that cannot be described using a local
operator product expansion has been recently carried out in
\cite{Benzke:2010js} (see also \cite{Lee:2006wn}). This analysis puts
the naive estimate of the size of non-local power corrections in
\cite{Misiak:2006zs, Misiak:2006ab} on firm theoretical grounds, and
at the same time indicates that a further reduction of the theoretical
uncertainty plaguing \eq{eq:SM} below $5 \%$ would require a
theoretical breakthrough.

Besides the branching ratio also the $\BXsga$ photon-energy spectrum
is of theoretical interest and phenomenological relevance
\cite{spectrum}. While close to the physical endpoint $E_\gamma =
m_B/2$ the spectrum is dominated by the $(Q_{1,2}, Q_7)$ and $(Q_7,
Q_7)$ contributions, the $(Q_8, Q_8)$ interference is numerically the
most important one for $E_\gamma \lesssim 1.1 \, \GeV$, because it
involves a soft singularity $1/E_\gamma$ related to photon
bremsstrahlung. The theoretical description of the $(Q_8, Q_8)$ part
of the spectrum has a simple, but important feature, that is
associated with the photon having a hadronic substructure, and
manifests itself in the appearance of collinear singularities in the
perturbative result of the fixed-order calculation.  The leading
contribution of the $(Q_8, Q_8)$ interference to the photon-energy
spectrum in $b \to X_s^{\rm part} \gamma$ has been known for some time
\cite{Ali:1995bi}. This contribution is suppressed by a single power
of $\as$ with the respect to the leading $(Q_7, Q_7)$ interference,
and therefore is part of the next-to-leading order (NLO) corrections
to the spectrum.

The $(Q_{1,2}, Q_{1,2})$, $(Q_{1,2}, Q_7)$, and $(Q_7, Q_8)$
corrections to the photon-energy spectrum were calculated in the
large-$\beta_0$ approximation, \ie, including terms of order $\beta_0
\hspace{0.15mm} \as^2$ through naive non-abelianization \cite{BLM},
already in \cite{Ligeti:1999ea}. However, in that work neither the
$(Q_{1,2}, Q_8)$ nor the ($Q_8, Q_8$) interference was considered. In
this article, we close this gap partly by calculating the corrections
to the photon-energy spectrum originating from the self-interference
of the chromomagnetic dipole operator $Q_8$ at $\ord (\beta_0
\hspace{0.15mm} \as^2)$. A calculation of the $(Q_{1,2}, Q_8)$
interference, that completes the $\ord (\as^2)$ calculation of the
spectrum in the large-$\beta_0$ approximation has recently been
completed and will soon be published \cite{Mikolaj}.

This article is organized as follows. In \Sec{sec:results} we provide
the analytic results of our calculation, while \Sec{sec:calculation}
contains a brief description of the calculation itself. The numerical
impact of the considered NNLO corrections on the branching ratio of
$\BXsga$ is studied in \Sec{sec:numerics}. We conclude in
\Sec{sec:conclusions}.

\section{Analytic Results}
\label{sec:results}

At the $\bar{B}$-meson mass scale $\mub = \ord (\mb)$ the partonic $b
\to X_s^{\rm part} \gamma$ cut rate can be expressed in terms of the
charmless semileptonic total decay width as
\begin{widetext}
\beq \label{eq:Gamma}
\Gamma (b \to X_s^{\rm part} \gamma)^{E_\gamma > E_0} = \f{6 \aem_{\rm
    em}}{\pi} \left | \f{V_{ts}^\ast V_{tb}}{V_{ub}} \right |^2 \Gamma
(b \to X_u^{\rm part} \ell \bar{\nu}) \sum_{i,j = 1}^8 C_i^{\rm eff}
(\mub) \hspace{0.25mm} C_j^{\rm eff} (\mub) \hspace{0.25mm} K_{ij}
(E_0) \, ,
\eeq
\end{widetext}
where $\aem_{\rm em} = \aem_{\rm em} (0) = 1/137.036$, $V_{kl}$ are
the relevant Cabibbo-Kobayashi-Maskawa matrix elements, and $C_i^{\rm
  eff} (\mub)$ denote the effective Wilson coefficients defined as in
\cite{Misiak:2006ab}.

In the following, we will present analytic formulas for the $\ord
(\beta_0 \hspace{0.15mm} \as^2)$ corrections to $K_{88} (E_0)$. This
function describes the self-interference of the chromomagnetic
dipole operator
\beq \label{eq:Q8}
Q_8 = \f{\gs}{16 \pi^2} \, \mb (\mu) \left ( \bar{s}_L \sigma^{\mu
    \nu} T^a b_R \right ) G^a_{\mu \nu} \, ,
\eeq  
where $\gs$ is the strong-coupling constant, $\mb (\mu)$ denotes the
running $\MSbar$ mass of the bottom quark, $q_{L,R}$ are left- and
right-chiral quark fields, $G_{\mu \nu}^a$ is the gluonic field
strength tensor, and $T^a$ are the color generators normalized to
${\rm Tr} \left (T^a T^b \right ) = \TR \delta^{ab}$ with $\TR = 1/2$.

Including QCD corrections up to NNLO, the coefficient $K_{88} (E_0)$
can be written (in a notation following closely the one adopted in
\cite{Misiak:2006ab}) as follows:
\beq \label{eq:K88exp}
K_{88} (E_0) = \sum_{n=1}^2 \left ( \f{\as(\mub)}{4 \pi} \right )^n
K_{88}^{(n)} (E_0) \, .
\eeq 

In agreement with \cite{Ali:1995bi, Kapustin:1995fk}, we find for
$K_{88}^{(1)} (E_0)$ the analytic expression
\begin{widetext}
\beq \label{eq:K881}
K_{88}^{(1)} (E_0) = \f{4}{27} \left \{ - \ln \frac{\mb^2}{\ms^2} \,
  \big [ \hspace{0.25mm} \delta (2 + \delta) + 4 \ln \bar \delta
  \hspace{0.25mm} \big ] + 4 \, {\rm Li}_2 \hspace{0.25mm} \bar \delta
  - \f{2\pi^2}{3} - \delta (2 + \delta) \ln \delta + 8 \ln \bar \delta
  -\f{2\delta^3 }{3} + 3 \delta^2 + 7 \delta \right \} .
\eeq
\end{widetext}
where $\delta = 1 - 2 E_0/\mb$ and $\bar \delta = 1 - \delta = 2
E_0/\mb$. As expected, the NLO function $K_{88}^{(1)} (E_0)$ is
logarithmically divergent for both $\bar \delta \to 0$ (soft
singularity) as well as $\ms \to 0$ (collinear singularity).  Notice
that terms suppressed by positive powers of the ratio $\ms/\mb$ have
been neglected in \eq{eq:K881}.

\begin{widetext}

\begin{figure}[!t]
\begin{center}
\mbox{\includegraphics[height=1in]{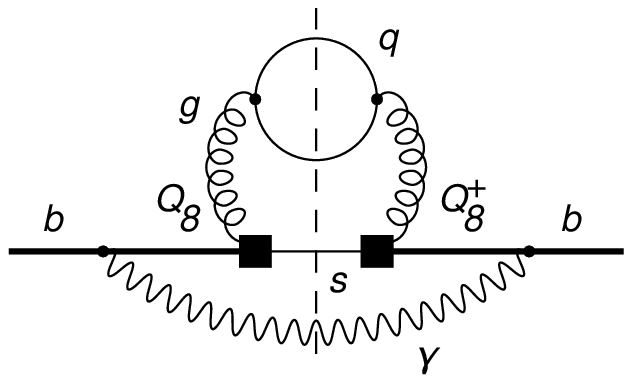}}
\qquad 
\mbox{\includegraphics[height=1in]{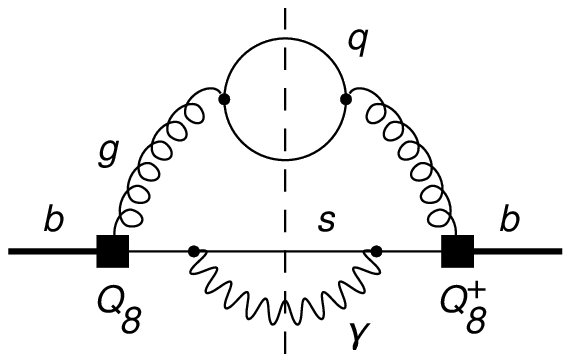}}
\qquad 
\mbox{\includegraphics[height=1in]{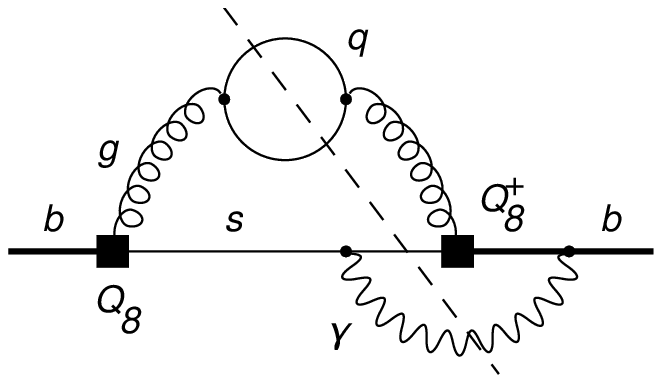}}
\vspace{0mm}
\caption{Four-particle cuts of the irreducible bottom-quark
  self-energy diagrams with a quark bubble contributing to the $b \to
  s \gamma q \bar q$ ($q = u,d,s$) transition at $\ord
  (\as^2)$. Left-right reflected diagrams are not shown. The second
  and third diagrams give rise to collinear logarithms $\ln \left
    (\mb^2/\ms^2 \right )$.}
\label{fig:K882NL}
\end{center}
\end{figure}

\end{widetext}

\begin{figure}[!t]
\begin{center}
\mbox{\includegraphics[height=1in]{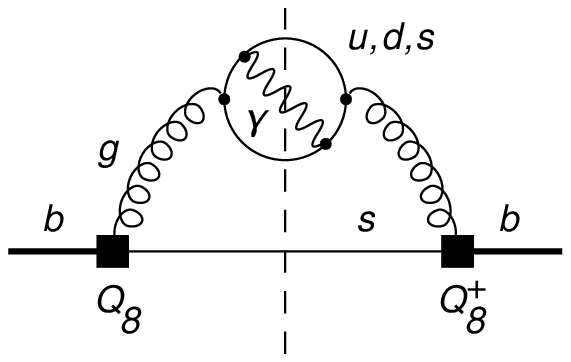}}

\vspace{3.5mm}

\mbox{\includegraphics[height=1in]{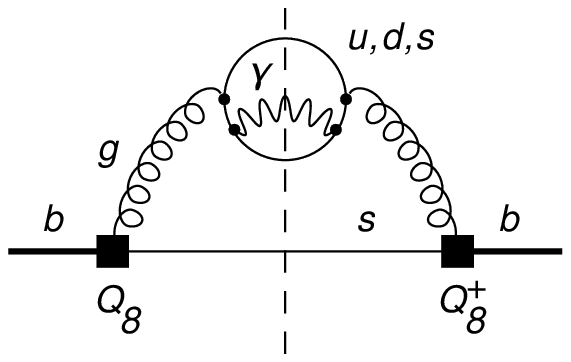}}
\vspace{0mm}
\caption{Four-particle cuts of the irreducible bottom-quark
  self-energy diagrams with a light-quark bubble contributing to the
  $b \to s \gamma u \bar{u}, d \bar{d}, s \bar{s}$ transition at $\ord
  (\as^2)$. Symmetric diagrams are not shown. The shown diagrams give
  rise to collinear logarithms $\ln \left (\mb^2/m_{u,d,s}^2 \right
  )$. In practice, these IR-sensitive terms are regulated by the
  light-meson masses $M_\pi$ and $M_K$.}
\label{fig:K882uds}
\end{center}
\end{figure}

\begin{widetext}

\begin{figure}[!t]
\begin{center}
\mbox{\includegraphics[height=1in]{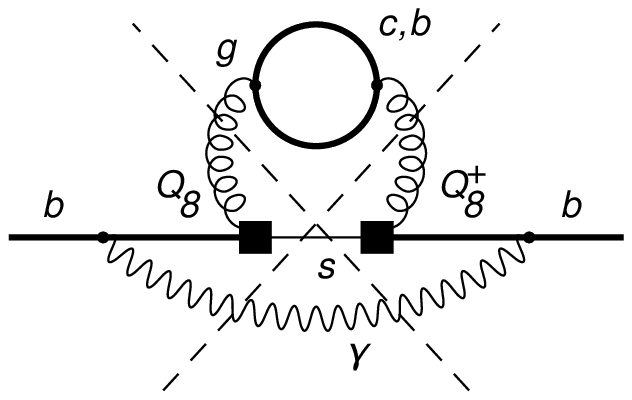}}
\qquad 
\mbox{\includegraphics[height=1in]{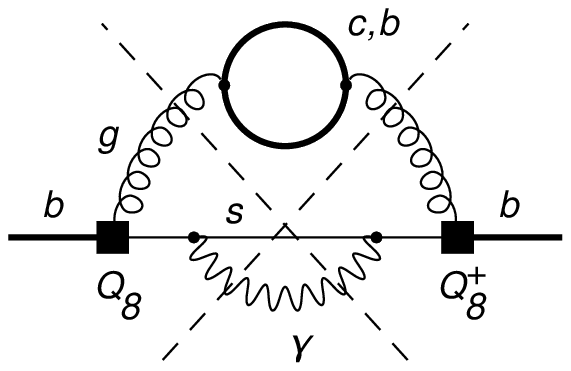}}
\qquad 
\vspace{3.5mm}
\mbox{\includegraphics[height=1in]{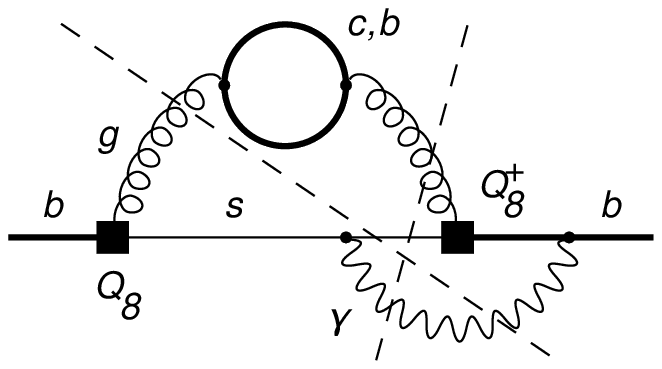}}
\vspace{0mm}
\caption{Three-particle cuts of the irreducible bottom-quark
  self-energy diagrams with a massive charm- and bottom-quark bubble
  contributing to the $b \to s \gamma g$ transition at $\ord
  (\as^2)$. Left-right reflected diagrams are not shown. The second
  and third diagrams give rise to collinear logarithms $\ln \left
    (\mb^2/\ms^2 \right )$.}
\label{fig:K882cb}
\end{center}
\end{figure}

\end{widetext}

The NNLO function $K_{88}^{(2)} (E_0)$ receives both fermionic and
purely gluonic contributions. The former corrections arise from the
Feynman diagrams shown in Figs.~\ref{fig:K882NL}, \ref{fig:K882uds},
and \ref{fig:K882cb}. Since in the large-$\beta_0$ approximation one
considers exclusively massless fermion-loop insertions in the gluon
propagators of the lower-order diagrams \cite{BLM}, it follows that in
this approximation only the light-quark ($q = u, d, s$) loop diagrams
shown in \Fig{fig:K882NL} need to be calculated. The graphs in
\Fig{fig:K882uds}, which also involve a light-quark bubble, belong to
a new channel which opens up at NNLO, and therefore are not captured
by the large-$\beta_0$ approximation. Also notice that massless quark
loops ($q = u, d, s$) in \Fig{fig:K882cb} involve a scaleless
integral, which implies that they evaluate to zero in dimensional
regularization. Consequently, we apply the naive non-abelianization
prescription to the light-quark diagrams in \Fig{fig:K882NL} only and
split $K_{88}^{(2)} (E_0)$ into a large-$\beta_0$ and a remaining part
\beq \label{eq:K882split}
  K_{88}^{(2)} (E_0) = K_{88}^{(2, \beta_0)} (E_0) + K_{88}^{(2,
    {\rm rem})} (E_0) \, , 
\eeq
with 
\beq \label{eq:K882beta0}
K_{88}^{(2, \beta_0)} (E_0) = -\f{3}{2} \, \beta_0 \hspace{0.25mm}
K_{88}^{(2, \NL)} (E_0) \, ,
\eeq
and $\beta_0 = 11 - 2/3 \, \left ( \NL + 2 \right )$.  As in the work
\cite{Misiak:2006ab}, we will set $\NL = 3$ in our numerical analysis.
Notice that effects related to the absence of real charm-quark pair
production in the partonic $b \to X_s^{\rm part} \gamma$ decay and to
non-zero values of the charm- and bottom-quark mass in quark bubbles
on the gluon propagators are by definition contained in $K_{88}^{(2,
  {\rm rem})} (E_0)$ and not in $K_{88}^{(2, \beta_0)} (E_0)$.

The coefficient $K_{88}^{(2, \NL)} (E_0)$ introduced in
\eq{eq:K882beta0} describes the contribution of the graphs in
\Fig{fig:K882NL} involving a single massless quark flavor. It can be
written as
\bea \label{eq:K882NL}
K_{88}^{(2, \NL)} (E_0) = \frac{4}{3} \, \TR \! \left [ \int_{\bar
    \delta}^1 \!  dz \hspace{0.2mm} F_{88}^{(2, \NL)} -K_{88}^{(1)}
  (E_0) L_b \right ] \! , \hspace{2.5mm}
\eea
where $L_b = \ln \left (\mub^2/\mb^2 \right )$, $z = 2
E_\gamma/\mb$, and 
\begin{widetext}
\beq \label{eq:F882NL}
\begin{split}
  F_{88} ^{(2, \NL)} = \f{8}{27} \, & \left \{ -\ln \f{\mb^2}{\ms^2}
    \left [ \, \f{36 - 41 z + 13 z^2 + 2 z^3}{6 z} - \f{1 + \bar
        z^2}{z} \ln \bar z \, \right ] + \f{1 + \bar z^2}{z} \left [
      \ln^2 \bar z - \f{\pi^2}{3} \right ] \right. \\ & \left.
    \phantom{c} - \f{60 - 65 z + 16 z^2 + 8 z^3}{6 z} \ln \bar z +
    \f{604 - 702 z + 126 z^2 + 107 z^3}{36 z} \right \} \, ,
\end{split}
\eeq 
\end{widetext}
with $\bar z = 1 - z$. The latter expression is the main analytic
result of our paper. Similar to \eq{eq:K881} also \eq{eq:F882NL}
contains a collinear divergence, which we have regulated by keeping a
non-vanishing strange-quark mass. Again terms suppressed by positive
powers of $\ms/\mb$ have been neglected in the function $F_{88} ^{(2,
  \NL)}$.

The function $K_{88}^{(2, {\rm rem})} (E_0)$ entering
(\ref{eq:K882split}) encodes the $\ord (\as^2)$ contributions to the
$(Q_8, Q_8)$ interference that are beyond the large-$\beta_0$
approximation. It takes the form
\begin{widetext}
\beq \label{eq:K882rem}
K_{88}^{(2, {\rm rem})} (E_0) = \f{33}{2} \, K_{88}^{(2, \NL)} (E_0) +
\sum_{q = u, d, s} K_{88}^{(2, q, \gamma)} (E_0) + \sum_{q = c, b}
K_{88}^{(2, q, M)} (E_0) + K_{88}^{(2, g)} (E_0) \,,
\eeq
\end{widetext}
The function $K_{88}^{(2,g)} (E_0)$ originates from diagrams with no
quark loops, while $K_{88}^{(2,q, \gamma)} (E_0)$ corresponds to
\Fig{fig:K882uds}. The calculation of these contributions is beyond
the scope of the present article. As was already mentioned, real $c
\bar c$ pair production is not included in $b \to X_s^{\rm part}
\gamma$ by definition, while $b \bar b$ pair production is
kinematically forbidden. Thus, $K_{88}^{(2,q,\gamma)} (E_0)$ is
non-vanishing for light quarks with $q=u,d,s$ only.

The function $K_{88}^{(2,q,M)} (E_0)$ originates from \Fig{fig:K882cb}
and vanishes for massless quarks ($q=u,d,s$) in dimensional
regularization due to the appearance of scaleless integrals. Its
analytic form for $q=c,b$ can be obtained by multiplying the NLO
coefficient $K_{88}^{(1)} (E_0)$ by a renormalized one-loop
vacuum-polarization function at zero-momentum transfer. Explicitly we
find
\beq \label{eq:K882i}
K_{88}^{(2, q)}(E_0) = - \frac{4}{3} \, \TR \hspace{0.25mm}
K_{88}^{(1)} (E_0) L_q \, ,
\eeq 
where $L_c = \ln \left (\mub^2/\mc^2 \right )$. Notice finally that
the effects of the diagrams in \Fig{fig:K882cb} can also been taken
into account through gluon wave-function renormalization in the NLO
graphs.

It is also straightforward to derive an expression for the function
$K_{88}^{(2, {\rm rem})} (E_0)$ in the large-$m_c$ limit.  The latter
enters the calculation of the $\BXsga$ branching ratio via an
interpolation in the charm-quark mass \cite{Misiak:2006ab}. In
agreement with that paper, we obtain the expression
\begin{widetext}
\beq \label{eq:K88b0rem}
  K_{88}^{(2, {\rm rem})} (E_0) = \left ( - \f{50}{3} +
    \f{8 \pi^2}{3}  - \f{2}{3} L_c \right ) K_{88}^{(1)} (E_0) +
  X_{88}^{(2, {\rm rem})} (E_0) \, .
\eeq 
\end{widetext}
Here the first term on the right-hand side is the leading term in the
large-$\mc$ expansion of $K_{88}^{(2, {\rm rem})} (E_0)$. It consists
of two parts, one arising from the normalization to the charmless
semileptonic rate and the other being proportional to $L_c$, which is
due to the $\MSbar$ matching corrections connecting the
strong-coupling constants of the effective four- and five-flavor
theories, as encoded in (\ref{eq:K882i}). The $m_c$-independent
quantity $X_{88}^{(2, {\rm rem})} (E_0)$ summarizes unknown $\ord
(\as^2)$ contributions arising from the self-interference $(Q_8, Q_8)$
in the theory with decoupled charm quark (together with the
corresponding bremsstrahlung).

\section{Calculational Technique}
\label{sec:calculation}

In order to calculate the $\ord (\beta_0 \hspace{0.15mm} \as^2)$
corrections to the partonic $b \to X_s^{\rm part} \gamma$ cut rate
arising from the self-interference $(Q_8, Q_8)$ we have employed the
optical theorem. In particular, we have exploited the one-to-one
correspondence between the interferences among diagrams contributing
to the process $b \to s \gamma q \bar{q}$ and the physical cuts of
three-loop bottom-quark self-energy diagrams. As can be seen by
glancing at \Fig{fig:K882NL}, we are interested in diagrams in which
the chromomagnetic dipole operator $Q_8$ appears on both sides of the
cut.  The contribution of a specific physical cut to the imaginary
parts of the corresponding bottom-quark self-energy diagrams is
thereby evaluated by means of the Cutkosky rules. See \cite{2ndQ7} for
more detailed discussions.

We have evaluated the relevant four-particle cuts in two different
ways.  First, by a direct computation of the light-quark contributions
using the set-up previously employed in the NNLO calculation of the
$(Q_7, Q_{7,8})$ contributions \cite{2ndQ7, Q7Q8}, and, second, by
performing the NLO calculation of the $(Q_8, Q_8)$ contribution with a
fictitious gluon mass which allows us to obtain the sought $\ord
(\as^2 )$ contributions from a dispersion integral over the gluon
virtuality \cite{Smith:1994id}.\footnote{This method was also used in
  the calculation of the $\ord (\beta_0 \hspace{0.15mm} \as^2)$
  corrections to the photon-energy spectrum of the $(Q_{1,2},
  Q_{1,2})$, $(Q_{1,2}, Q_7)$, and $(Q_7, Q_8)$ terms
  \cite{Ligeti:1999ea}. We verified the correctness of the $(Q_7,
  Q_8)$ contribution given in the aforementioned article.}  For a
recent detailed review of this technique we refer to
\cite{Benson:2004sg}. In both cases, the reduction to master integrals
via the Laporta algorithm \cite{Laporta:2001dd} has been carried out
keeping a non-vanishing strange-quark mass to regulate the residual
collinear divergences. All the other infrared (IR) or ultraviolet
divergences, appearing in intermediate stages of the calculation, have
been regulated dimensionally in $d = 4 - 2 \, \eps$ dimensions. The
master integrals have been evaluated analytically both by direct
integration over the phase space and by employing the differential
equation method \cite{Remiddi:1997ny}. Throughout the calculation of
the master integrals, we have neglected terms suppressed by positive
powers of the ratio $\ms/\mb$. The agreement of the results obtained
by the two methods serves as a powerful check of our calculation.

\section{Numerical Analysis}
\label{sec:numerics}

In the following, we will investigate the numerical size of the $\ord
(\as^2)$ contributions to the $(Q_8, Q_8)$ interference at the level
of the branching ratio of $\BXsga$. In order to simplify the
comparison with the existing literature, we will adopt the
conventions, the notations, and the numerical input parameters
employed in \cite{Misiak:2006ab}. Specifically, we will use $\mub =
2.5 \, {\rm GeV}$, $\mb = 4.68 \, {\rm GeV}$, $\mb/\ms = 50$,
$\BRXc_{\rm exp} = 10.61 \%$, $C = 0.58$, $|V_{ts}^\ast
V_{tb}/V_{cb}|^2 = 0.9676$, $\as (2.5 \, {\rm GeV}) = 0.271$, and
$C_8^{{\rm eff} (0)} (2.5 \, {\rm GeV} ) = -0.171$. With this choice
of input, one exactly reproduces the central value of the SM
prediction \eq{eq:SM}.

\begin{widetext}

\begin{figure}[!t]
\begin{center}
\mbox{\includegraphics[height=2in]{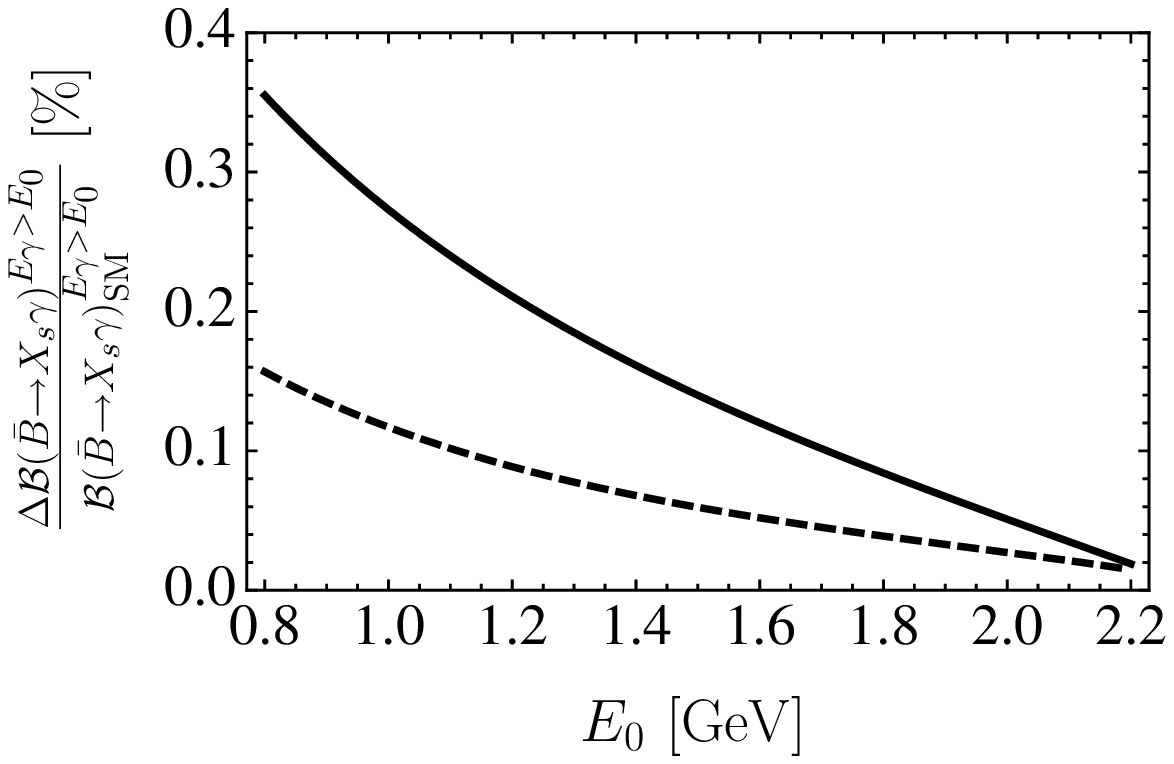}}  
\qquad 
\mbox{\includegraphics[height=2in]{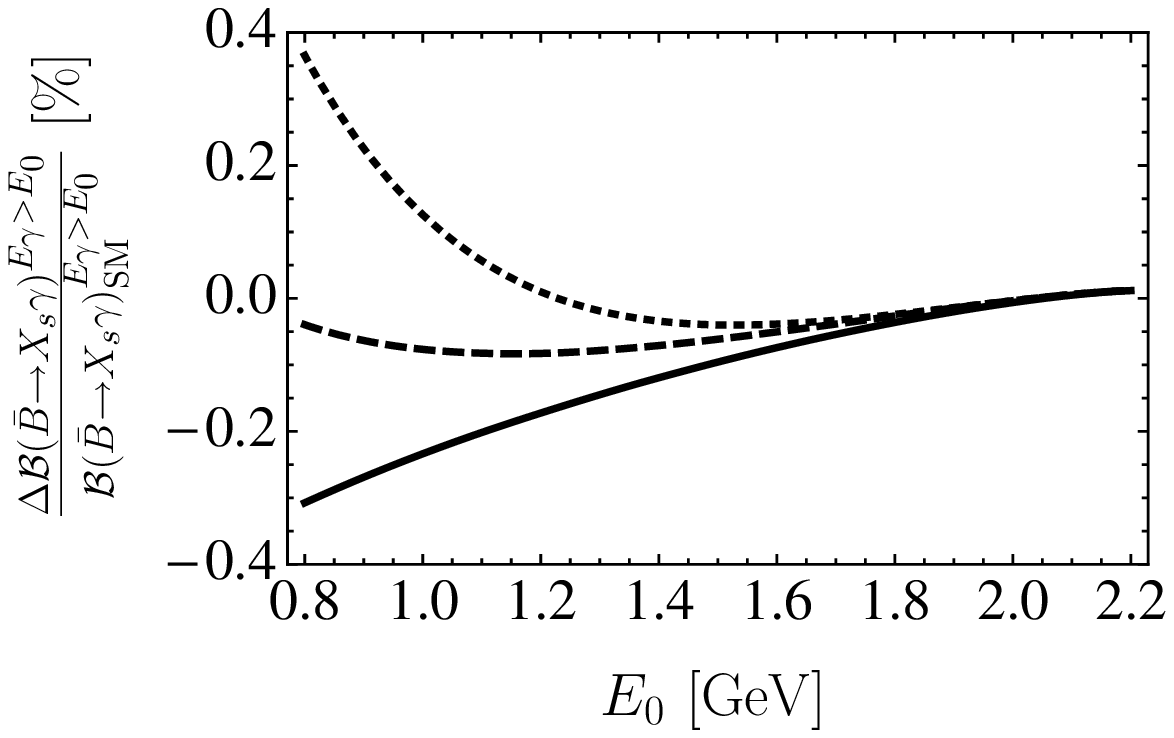}}
\vspace{1mm}
\caption{Left panel: Relative changes in $\BRga_{\rm SM}^{E_\gamma >
    E_0}$ due to the $\ord (\beta_0 \hspace{0.15mm} \as^2)$
  contributions to $(Q_8, Q_8)$. The solid (dashed) line shows the
  fixed-order (resummed) result as a function of the photon-energy cut
  $E_0$. Right panel: Comparison of perturbative and non-perturbative
  corrections related to the self-interference $(Q_8, Q_8)$. The solid
  line indicates the relative shift in $\BRga_{\rm SM}^{E_\gamma >
    E_0}$ due to a resummation of collinear effects, while the dashed
  and dotted lines illustrate the impact of the non-perturbative
  components of the photon-fragmentation functions assuming two
  different models of vector-meson dominance. See text for further
  details. }
\label{fig:numerics}
\end{center}
\end{figure}

\end{widetext}

We start by considering the impact of the large-$\beta_0$
corrections. In this limit, we can write the correction to the
$\BXsga$ branching ratio arising from the $(Q_8, Q_8)$ interference at
$\ord (\as^2)$ as
\begin{widetext}
\beq \label{eq:b0num}
\Delta {\cal B} (\BXsga)^{E_\gamma > E_0} = \BRXc_{\rm exp} \, \f{6
  \aem_{\rm em}}{\pi C} \left | \f{V_{ts}^\ast V_{tb}}{V_{cb}} \right
|^2 \, \big |C_8^{{\rm eff} (0)} (\mu_b) \big |^2 \, \left
  (\frac{\as (\mub)}{4 \pi} \right )^2 \, K_{88}^{(2,\beta_0)} (E_0) \,,  
\eeq
\end{widetext}
here $C$ is the so-called semileptonic phase-space factor and
$K_{88}^{(2,\beta_0)} (E_0)$ has already been defined in
\eq{eq:K882beta0}.  In the left panel of \Fig{fig:numerics} we show
$\Delta {\cal B} (\BXsga)^{E_\gamma > E_0}$ normalized to the central
value of ${\cal B} (\BXsga)_{\rm SM}^{E_\gamma > E_0}$ as a function
of the photon-energy cut $E_0$. We see from the solid line that the
inclusion of the $\ord (\beta_0 \hspace{0.15mm} \as^2)$ contributions
leads to a relative change of the $\BXsga$ branching ratio of $+0.12
\%$ ($+0.27 \%$) for $E_0 = 1.6 \, \GeV$ ($E_0 = 1.0 \, \GeV$). We
recall that for the two chosen values of $E_0$, the shifts due to the
NLO corrections involving $(Q_8, Q_8)$ amount to $+0.24\%$ and
$+0.66\%$, respectively. These numbers imply that after naive
non-abelianization the term (\ref{eq:K882NL}) constitutes a correction
of almost $50\%$ with respect to the $\ord ( \as)$ contributions.

As we have already mentioned, in the $(Q_8, Q_8)$ interference also
the corrections which are not part of the large-$\beta_0$
approximation (such as the four-particle cut diagrams in
\Fig{fig:K882uds}) involve collinear logarithms associated with photon
fragmentation of $b \to s g$. Sufficiently far away from the endpoint
of the photon-energy spectrum, the resulting IR-sensitive terms can be
subtracted and absorbed into non-perturbative photon-fragmentation
functions \cite{Kapustin:1995fk},\footnote{In fact, in
  \cite{Kapustin:1995fk} only purely perturbative corrections are
  included.} which obey perturbative evolution
(Dokshitzer-Gribov-Lipatov-Altarisi-Parisi or DGLAP) equations with
non-perturbative initial distributions to be extracted from
experiment.\footnote{In the endpoint region the non-perturbative
  physics associated with the $(Q_8, Q_8)$ interference is encoded in
  a complicated subleading four-quark shape function rather than a
  fragmentation function. A detailed discussion of how these effects
  factorize has been given in \cite{Benzke:2010js} using
  soft-collinear effective theory.} While a complete calculation of
collinear effects at $\ord (\as^2)$ is beyond the scope of the present
article, we find it illustrative to study the issue of IR-sensitive
contributions arising in \eq{eq:K881}, \eq{eq:K882NL}, and
\eq{eq:F882NL}. From this exercise we expect to get an idea about the
potential size of both resummation and non-perturbative effects
associated with the $(Q_8, Q_8)$ self-interference.

The resummation of the collinear logarithms appearing in the
$K_{88}^{(2,\NL)} (E_0)$ corrections is achieved by convoluting the
hard function $C_s^{\NL} (x)$, that describes the process $b \to s q
\bar q$ for fixed energy $x$ of the strange quark, with the universal
strange-quark-to-photon fragmentation function $D_{s \to \gamma}
(x)$. Explicitly, we find that the result of the resummation of the
collinear logarithm in \eq{eq:F882NL} takes the form\footnote{In the
  absence of QCD, the expression for the strange-quark-to-photon
  fragmentation function is given by $D_{s \to \gamma} (x) = \aem_{\rm
    em} \hspace{0.25mm} Q_s^2/(2 \pi) \, \big (1 + \bar x^2 \big )/x
  \, \ln (\mub^2/\mu_s^2)$ with $Q_s = -1/3$, $\mub \approx \mb$, and
  $\mu_s \approx \ms \approx \Lambda_{\rm QCD}$. Substituting this
  result into \eq{eq:F882NLtilde}, one recovers the terms in
  \eq{eq:F882NL} that are singular in the limit $\ms \to 0$.}
\beq \label{eq:F882NLtilde} 
\tilde F_{88}^{(2,\NL)} = \frac{2 \pi}{\aem_{\rm em}} \int_z^1 \!
\frac{dx}{x} \, C_s^{\NL}(x) \, D_{s \to \gamma} \left ( \frac{z}{x}
\right ) ,
\eeq 
with
\beq \label{eq:Cs} 
C_s^{\NL} (x) = -\frac{8}{3} \left( \frac{10}{3} \, \delta (\bar x) -
  \left [ \frac{1}{\bar x} \right ]_+ + 1 + x -\frac{x^2}{2} \right) ,
\eeq
where $\bar x = 1 -x$ and $\left [ 1/\bar x \right ]_+$ denotes the
usual plus distribution. A factorization formula similar to the one
given in \eq{eq:F882NLtilde} can also be derived for the complete
$\ord (\as^2)$ correction to $K_{88} (E_0)$ in the collinear limit.

The full photon-fragmentation functions $D_{i \to \gamma} (x)$ with $i
= s, g$ are sums of perturbative and non-perturbative
components. While the former are fully calculable in QCD, the latter
have to be modeled. Following \cite{Bourhis:1997yu}, which the
interested reader should consult for further details, we will employ a
vector-meson dominance model and assume that quarks and gluons first
fragment into vector mesons which then turn into photons. We begin our
discussion by studying the impact of the anomalous parts of the
photon-fragmentation functions, \ie, the components encoding the
perturbative evolution as described by the inhomogeneous DGLAP
equations.  Comparing the resummed with the fixed-order $\ord (\beta_0
\hspace{0.15mm} \as^2)$ result, as indicated by the dashed and solid
lines in the left panel of \Fig{fig:numerics}, respectively, we infer
that the resummation of collinear logarithms decreases the obtained
results. Numerically, we find a relative change of $+0.05 \%$ ($+0.12
\%$) for $E_0 = 1.6 \, \GeV$ ($E_0 = 1.0 \, \GeV$), which implies that
the resummation suppresses the considered correction by more than a
factor of 2. We also mention that for photon-energy cuts around $1.6
\, {\rm GeV}$ the resummation of collinear logarithms appearing in the
${\cal O} (\beta_0 \hspace{0.15mm} \as^2)$ correction can be
effectively described by choosing $\mb/\ms = 14$ in the analytic
expression (\ref{eq:F882NL}).

We now turn our attention to the non-perturbative contributions
related to the photon fragmentation from $b \to s g$. These
corrections turn out to be potentially larger than the resummation
effects. This is illustrated by the right panel in \Fig{fig:numerics},
which displays the relative change in $\BRga_{\rm SM}^{E_\gamma >
  E_0}$ arising from the sum of the $\ord (\as)$ and $\ord (\beta_0
\hspace{0.15mm} \as^2)$ corrections to $(Q_8, Q_8)$, including the
anomalous parts of $D_{i \to \gamma} (x)$ only (solid line) and
employing the full photon-fragmentation functions with two different
non-perturbative initial conditions (dashed and dotted lines). In each
case, we have subtracted the fixed-order $\ord (\as)$ corrections
\eq{eq:K881} from our results, since these effects are already part of
the SM prediction \eq{eq:SM}.  We see again that the choice $\mb/\ms =
50$, adopted throughout the recent literature on $\BXsga$, tends to
overestimate the effects of resumming collinear
logarithms. Numerically, we find relative shifts of $-0.07\%$ and
$-0.23\%$ for $E_0 = 1.6 \, {\rm GeV}$ and $E_0 = 1.0 \, {\rm GeV}$,
respectively. After incorporating on top of the anomalous also the
non-perturbative components of $D_{i \to \gamma} (x)$, we obtain
instead corrections of $-0.05\%$ and $-0.04\%$ or $-0.04\%$ and
$0.37\%$. The former (latter) numbers correspond to set I (II) of the
full photon-fragmentation functions $D_{i \to \gamma} (x)$ determined
in \cite{Bourhis:1997yu}. We recall that while the initial conditions
of the quark-to-photon fragmentation functions are well constrained by
$e^+ e^-$ data, the one of the gluon-to-photon fragmentation function
is not. Compared to set I, the gluon-to-photon fragmentation function
of set II is significantly larger, in particular, for small $x$.
Since the function $D_{g \to \gamma} (x)$ enters the resummation of
collinear logarithms in $(Q_8, Q_8)$ at $\ord (\as)$ via
\cite{Kapustin:1995fk}
\beq \label{eq:K881resum} 
\tilde K_{88}^{(1)} (E_0) = \frac{2 \pi}{\aem_{\rm em}} \int_{\bar
  \delta}^1 \! dz \; \frac{8}{3} \, \big [ D_{s \to \gamma} (z) + D_{g
  \to \gamma} (z) \big ] \,,
\eeq 
this results in larger shifts for set II than for set I.

In conclusion, our study of non-perturbative effects related to photon
fragmentation seems to indicate, first, that the size of hadronic
effects associated to the interference of $(Q_8, Q_8)$ should not
shift the central value of \eq{eq:SM} by more than $+1\%$ and, second,
that setting $\mb/\ms = 50$ in the terms $\ln \left (\mb^2/\ms^2
\right )$ entering the fixed-order result allows one to capture most
of the numerical effect. A recent much more detailed study
\cite{Benzke:2010js} finds slightly larger non-perturbative effects of
$[-0.3, +1.9]\%$ related to the self-interference of the
chromomagnetic dipole operator $Q_8$. While a straightforward
comparison of this result with ours is difficult, given the very
different nature of the used framework, the fact that the two
calculations result in numbers in the same ballpark gives us further
confidence that hadronic contributions in $\BXsga$ related to $(Q_8,
Q_8)$ indeed represent a minor effect.

\section{Conclusions}
\label{sec:conclusions}

In this work we have calculated the NNLO corrections to the $b \to
X_s^{\rm part} \gamma$ photon-energy spectrum in the large-$\beta_0$
approximation that arise from self-interference contribution of the
chromomagnetic dipole operator $Q_8$. The contributions from $(Q_8,
Q_8)$ are known to be numerically subleading at NLO for the
photon-energy cut currently employed in the measurements of the
$\BXsga$ branching ratio. We find that this trend continues at NNLO
and that the calculated $\ord (\beta_0 \hspace{0.15mm} \as^2)$
corrections have only a marginal impact on the $\BXsga$ branching
ratio, amounting to a relative shift of a few permille. However,
corrections to the spectrum arising from the $(Q_8,Q_8)$ interference
are theoretically interesting in their own right, since they are known
to be logarithmically divergent in the limit of vanishing photon
energy, and because they contain collinear singularities that are
associated with the intrinsic hadronic component of the
photon. Concerning the latter issue, we have shown that
non-perturbative effects in $(Q_8, Q_8)$ due to photon fragmentation
from $b \to s g$ presumably constitute an effect of below a percent
only. Our results can be readily incorporated in the SM calculation of
the $\BXsga$ branching ratio. While a total non-perturbative
uncertainty of about $5 \%$ will affect the SM prediction for the
branching ratio for some time to come, it is still mandatory to update
the available NNLO estimate by including all the $\ord (\as^2)$
corrections which were calculated in the past four years, with the aim
of reducing as much as possible the residual perturbative
uncertainty. The calculation presented here, constitutes a necessary
ingredient to achieve this goal.

\subsubsection*{Acknowledgments}
\label{subsec:acknowledgments}

We are grateful to Miko{\l}aj Misiak and Michal Poradzi{\'n}ski for
verifying the numerical impact of our ${\cal O} (\beta_0
\hspace{0.15mm} \as^2)$ result (see also \cite{Mikolaj}). We thank
Miko{\l}aj Misiak for his valuable comments on the manuscript, which
allowed us to improve it, and useful discussions. We also thank
Michael Benzke and Matthias Neubert for reading an almost final
version of our article and providing their suggestions. Helpful
conversations with Babis Anastasiou, Michael Benzke, Thomas Gehrmann,
Sebastian J\"ager, Matthias Neubert, and Giulia Zanderighi are
acknowledged. This work made use of {\tt AIR}
\cite{Anastasiou:2004vj}, {\tt FORM} \cite{Vermaseren:2000nd}, {\tt
  HPL} \cite{HPL}, {\tt HypExp} \cite{HypExp}, and {\tt JaxoDraw}
\cite{Binosi:2003yf}, and has been supported in part by the Schweizer
Nationalfonds and the European Organization for Nuclear Research.

\end{document}